%% file: root.tex
\documentclass{ifacconf}

\usepackage{graphicx}      % include this line if your document contains figures

\makeatletter
\let\old@ssect\@ssect % Store how ifacconf defines \@ssect
\makeatother
\usepackage{natbib}        % required for bibliography
\bibstyle{ifacconf}
\usepackage[hidelinks]{hyperref}
\makeatletter
\def\@ssect#1#2#3#4#5#6{%
  \NR@gettitle{#6}% Insert key \nameref title grab
  \old@ssect{#1}{#2}{#3}{#4}{#5}{#6}% Restore ifacconf's \@ssect
}
\makeatother

\usepackage{amsmath}
\usepackage{amsfonts}
\usepackage{amssymb}

\usepackage{caption}
\usepackage{subcaption}
\usepackage{tabularx}

\usepackage{booktabs}
\usepackage{enumitem}

\usepackage{xcolor}
\usepackage{listings}
\lstdefinelanguage{Julia}{
    keywords=[3]{abs,abs2,abspath,accept,accumulate,accumulate!,acos,acos_fast,acosd,acosh,acosh_fast,acot,acotd,acoth,acsc,acscd,acsch,adjoint,adjoint!,all,all!,allunique,angle,angle_fast,any,any!,append!,apropos,ascii,asec,asecd,asech,asin,asin_fast,asind,asinh,asinh_fast,assert,asyncmap,asyncmap!,atan,atan2,atan2_fast,atan_fast,atand,atanh,atanh_fast,atexit,atreplinit,axes,backtrace,base,basename,beta,big,bin,bind,binomial,bitbroadcast,bitrand,bits,bitstring,bkfact,bkfact!,blkdiag,broadcast,broadcast!,broadcast_getindex,broadcast_setindex!,bswap,bytes2hex,cat,catch_backtrace,catch_stacktrace,cbrt,cbrt_fast,cd,ceil,cfunction,cglobal,charwidth,checkbounds,checkindex,chmod,chol,cholfact,cholfact!,chomp,chop,chown,chr2ind,circcopy!,circshift,circshift!,cis,cis_fast,clamp,clamp!,cld,clipboard,close,cmp,coalesce,code_llvm,code_lowered,code_native,code_typed,code_warntype,codeunit,codeunits,collect,colon,complex,cond,condskeel,conj,conj!,connect,consume,contains,convert,copy,copy!,copysign,copyto!,cor,cos,cos_fast,cosc,cosd,cosh,cosh_fast,cospi,cot,cotd,coth,count,count_ones,count_zeros,countlines,countnz,cov,cp,cross,csc,cscd,csch,ctime,ctranspose,ctranspose!,cummax,cummin,cumprod,cumprod!,cumsum,cumsum!,current_module,current_task,dec,deepcopy,deg2rad,delete!,deleteat!,den,denominator,deserialize,det,detach,diag,diagind,diagm,diff,digits,digits!,dirname,disable_sigint,display,displayable,displaysize,div,divrem,done,dot,download,dropzeros,dropzeros!,dump,eachcol,eachindex,eachline,eachmatch,edit,eig,eigfact,eigfact!,eigmax,eigmin,eigvals,eigvals!,eigvecs,eltype,empty,empty!,endof,endswith,enumerate,eof,eps,equalto,error,esc,escape_string,evalfile,exit,exp,exp10,exp10_fast,exp2,exp2_fast,exp_fast,expanduser,expm,expm!,expm1,expm1_fast,exponent,extrema,eye,factorial,factorize,falses,fd,fdio,fetch,fieldcount,fieldname,fieldnames,fieldoffset,filemode,filesize,fill,fill!,filter,filter!,finalize,finalizer,find,findfirst,findin,findlast,findmax,findmax!,findmin,findmin!,findn,findnext,findnz,findprev,first,fld,fld1,fldmod,fldmod1,flipbits!,flipdim,flipsign,float,floor,flush,fma,foldl,foldr,foreach,frexp,full,fullname,functionloc,gamma,gc,gc_enable,gcd,gcdx,gensym,get,get!,get_zero_subnormals,getaddrinfo,getalladdrinfo,gethostname,getindex,getipaddr,getkey,getnameinfo,getpeername,getpid,getsockname,givens,gperm,gradient,hash,haskey,hcat,hessfact,hessfact!,hex,hex2bytes,hex2bytes!,hex2num,homedir,htol,hton,hvcat,hypot,hypot_fast,identity,ifelse,ignorestatus,im,imag,include_dependency,include_string,ind2chr,ind2sub,indexin,indices,indmax,indmin,info,insert!,instances,intersect,intersect!,inv,invmod,invperm,invpermute!,ipermute!,ipermutedims,is,is_apple,is_bsd,is_linux,is_unix,is_windows,isabspath,isapprox,isascii,isassigned,isbits,isblockdev,ischardev,isconcrete,isconst,isdiag,isdir,isdirpath,isempty,isequal,iseven,isfifo,isfile,isfinite,ishermitian,isimag,isimmutable,isinf,isinteger,isinteractive,isleaftype,isless,isletter,islink,islocked,ismarked,ismatch,ismissing,ismount,isnan,isodd,isone,isopen,ispath,isperm,isposdef,isposdef!,ispow2,isqrt,isreadable,isreadonly,isready,isreal,issetgid,issetuid,issocket,issorted,issparse,issticky,issubnormal,issubset,issubtype,issymmetric,istaskdone,istaskstarted,istextmime,istril,istriu,isvalid,iswritable,iszero,join,joinpath,keys,keytype,kill,kron,last,lbeta,lcm,ldexp,ldltfact,ldltfact!,leading_ones,leading_zeros,length,less,lexcmp,lexless,lfact,lgamma,lgamma_fast,linearindices,linreg,linspace,listen,listenany,lock,log,log10,log10_fast,log1p,log1p_fast,log2,log2_fast,log_fast,logabsdet,logdet,logging,logm,logspace,lpad,lq,lqfact,lqfact!,lstat,lstrip,ltoh,lu,lufact,lufact!,lyap,macroexpand,map,map!,mapfoldl,mapfoldr,mapreduce,mapreducedim,mapslices,mark,match,matchall,max,max_fast,maxabs,maximum,maximum!,maxintfloat,mean,mean!,median,median!,merge,merge!,method_exists,methods,methodswith,middle,midpoints,mimewritable,min,min_fast,minabs,minimum,minimum!,minmax,minmax_fast,missing,mkdir,mkpath,mktemp,mktempdir,mod,mod1,mod2pi,modf,module_name,module_parent,mtime,muladd,mv,names,nb_available,ncodeunits,ndigits,ndims,next,nextfloat,nextind,nextpow,nextpow2,nextprod,nnz,nonzeros,norm,normalize,normalize!,normpath,notify,ntoh,ntuple,nullspace,num,num2hex,numerator,nzrange,object_id,occursin,oct,oftype,one,ones,oneunit,open,operm,ordschur,ordschur!,pairs,parent,parentindexes,parentindices,parse,partialsort,partialsort!,partialsortperm,partialsortperm!,peakflops,permute,permute!,permutedims,permutedims!,pi,pinv,pipeline,pointer,pointer_from_objref,pop!,popdisplay,popfirst!,position,pow_fast,powermod,precision,precompile,prepend!,prevfloat,prevind,prevpow,prevpow2,print,print_shortest,print_with_color,println,process_exited,process_running,prod,prod!,produce,promote,promote_rule,promote_shape,promote_type,push!,pushdisplay,pushfirst!,put!,pwd,qr,qrfact,qrfact!,quantile,quantile!,quit,rad2deg,rand,rand!,randcycle,randcycle!,randexp,randexp!,randjump,randn,randn!,randperm,randperm!,randstring,randsubseq,randsubseq!,range,rank,rationalize,read,read!,readandwrite,readavailable,readbytes!,readchomp,readdir,readline,readlines,readlink,readstring,readuntil,real,realmax,realmin,realpath,recv,recvfrom,redirect_stderr,redirect_stdin,redirect_stdout,redisplay,reduce,reducedim,reenable_sigint,reim,reinterpret,reload,relpath,rem2pi,repeat,replace,replace!,repmat,repr,reprmime,reset,reshape,resize!,rethrow,retry,reverse,reverse!,reverseind,rm,rol,rol!,ror,ror!,rot180,rotl90,rotr90,round,rounding,rowvals,rpad,rsearch,rsearchindex,rsplit,rstrip,run,scale!,schedule,schur,schurfact,schurfact!,search,searchindex,searchsorted,searchsortedfirst,searchsortedlast,sec,secd,sech,seek,seekend,seekstart,select,select!,selectperm,selectperm!,send,serialize,set_zero_subnormals,setdiff,setdiff!,setenv,setindex!,setprecision,setrounding,shift!,show,showall,showcompact,showerror,shuffle,shuffle!,sign,signbit,signed,signif,significand,similar,sin,sin_fast,sinc,sincos,sind,sinh,sinh_fast,sinpi,size,sizehint!,sizeof,skip,skipchars,skipmissing,sleep,slicedim,sort,sort!,sortcols,sortperm,sortperm!,sortrows,sparse,sparsevec,spawn,spdiagm,speye,splice!,split,splitdir,splitdrive,splitext,spones,sprand,sprandn,sprint,spzeros,sqrt,sqrt_fast,sqrtm,squeeze,srand,stacktrace,start,startswith,stat,std,stdm,step,stride,strides,string,stringmime,strip,strwidth,sub2ind,subtypes,success,sum,sum!,sumabs,sumabs2,summary,supertype,svd,svdfact,svdfact!,svdvals,svdvals!,sylvester,symdiff,symdiff!,symlink,systemerror,take!,takebuf_array,takebuf_string,tan,tan_fast,tand,tanh,tanh_fast,task_local_storage,tempdir,tempname,thisind,tic,time,time_ns,timedwait,to_indices,toc,toq,touch,trace,trailing_ones,trailing_zeros,transcode,transpose,transpose!,tril,tril!,triu,triu!,trues,trunc,truncate,trylock,tryparse,typeintersect,typejoin,typemax,typemin,unescape_string,union,union!,unique,unique!,unlock,unmark,unsafe_copy!,unsafe_copyto!,unsafe_load,unsafe_pointer_to_objref,unsafe_read,unsafe_store!,unsafe_string,unsafe_trunc,unsafe_wrap,unsafe_write,unshift!,unsigned,uperm,valtype,values,var,varinfo,varm,vcat,vec,vecdot,vecnorm,versioninfo,view,wait,walkdir,warn,which,whos,widemul,widen,withenv,workspace,write,xor,yield,yieldto,zero,zeros,zip,applicable,eval,fieldtype,getfield,invoke,isa,isdefined,nfields,nothing,setfield!,throw,tuple,typeassert,typeof,uninitialized,undef},%
    keywords=[3]{asum,axpby!,axpy!,blascopy!,dot,dotc,dotu,gbmv,gbmv!,gemm,gemm!,gemv,gemv!,ger!,hemm,hemm!,hemv,hemv!,her!,her2k,her2k!,herk,herk!,iamax,nrm2,sbmv,sbmv!,scal,scal!,symm,symm!,symv,symv!,syr!,syr2k,syr2k!,syrk,syrk!,trmm,trmm!,trmv,trmv!,trsm,trsm!,trsv,trsv!),abs,abs2,abspath,accept,accumulate,accumulate!,acos,acos_fast,acosd,acosh,acosh_fast,acot,acotd,acoth,acsc,acscd,acsch,adjoint,adjoint!,all,all!,allunique,angle,angle_fast,any,any!,append!,apropos,argmax,argmin,ascii,asec,asecd,asech,asin,asin_fast,asind,asinh,asinh_fast,assert,asyncmap,asyncmap!,atan,atan2,atan2_fast,atan_fast,atand,atanh,atanh_fast,atexit,atreplinit,axes,backtrace,base,basename,beta,bfft,bfft!,big,bin,bind,binomial,bitbroadcast,bitrand,bits,bitstring,bkfact,bkfact!,blkdiag,brfft,broadcast,broadcast!,broadcast_getindex,broadcast_setindex!,bswap,bytes2hex,cat,catch_backtrace,catch_stacktrace,cbrt,cbrt_fast,cd,ceil,cfunction,cglobal,charwidth,checkbounds,checkindex,chmod,chol,cholfact,cholfact!,chomp,chop,chown,chr2ind,circcopy!,circshift,circshift!,cis,cis_fast,clamp,clamp!,cld,clipboard,close,cmp,coalesce,code_llvm,code_lowered,code_native,code_typed,code_warntype,codeunit,codeunits,collect,colon,complex,cond,condskeel,conj,conj!,connect,consume,contains,conv,conv2,convert,copy,copy!,copysign,copyto!,cor,cos,cos_fast,cosc,cosd,cosh,cosh_fast,cospi,cot,cotd,coth,count,count_ones,count_zeros,countlines,countnz,cov,cp,cross,csc,cscd,csch,ctime,ctranspose,ctranspose!,cummax,cummin,cumprod,cumprod!,cumsum,cumsum!,current_module,current_task,dct,dct!,dec,deconv,deepcopy,deg2rad,delete!,deleteat!,den,denominator,deserialize,det,detach,diag,diagind,diagm,diff,digits,digits!,dirname,disable_sigint,display,displayable,displaysize,div,divrem,done,dot,download,dropzeros,dropzeros!,dump,eachcol,eachindex,eachline,eachmatch,edit,eig,eigfact,eigfact!,eigmax,eigmin,eigvals,eigvals!,eigvecs,eltype,empty,empty!,endof,endswith,enumerate,eof,eps,equalto,error,esc,escape_string,evalfile,exit,exp,exp10,exp10_fast,exp2,exp2_fast,exp_fast,expand,expanduser,expm,expm!,expm1,expm1_fast,exponent,extrema,eye,factorial,factorize,falses,fd,fdio,fetch,fft,fft!,fftshift,fieldcount,fieldname,fieldnames,fieldoffset,filemode,filesize,fill,fill!,filt,filt!,filter,filter!,finalize,finalizer,find,findfirst,findin,findlast,findmax,findmax!,findmin,findmin!,findn,findnext,findnz,findprev,first,fld,fld1,fldmod,fldmod1,flipbits!,flipdim,flipsign,float,floor,flush,fma,foldl,foldr,foreach,frexp,full,fullname,functionloc,gamma,gc,gc_enable,gcd,gcdx,gensym,get,get!,get_zero_subnormals,getaddrinfo,getalladdrinfo,gethostname,getindex,getipaddr,getkey,getnameinfo,getpeername,getpid,getsockname,givens,gperm,gradient,hash,haskey,hcat,hessfact,hessfact!,hex,hex2bytes,hex2bytes!,hex2num,homedir,htol,hton,hvcat,hypot,hypot_fast,idct,idct!,identity,ifelse,ifft,ifft!,ifftshift,ignorestatus,im,imag,include_dependency,include_string,ind2chr,ind2sub,indexin,indices,indmax,indmin,info,insert!,instances,intersect,intersect!,inv,invmod,invperm,invpermute!,ipermute!,ipermutedims,irfft,is,is_apple,is_bsd,is_linux,is_unix,is_windows,isabspath,isapprox,isascii,isassigned,isbits,isblockdev,ischardev,isconcrete,isconst,isdiag,isdir,isdirpath,isempty,isequal,iseven,isfifo,isfile,isfinite,ishermitian,isimag,isimmutable,isinf,isinteger,isinteractive,isleaftype,isless,isletter,islink,islocked,ismarked,ismatch,ismissing,ismount,isnan,isodd,isone,isopen,ispath,isperm,isposdef,isposdef!,ispow2,isqrt,isreadable,isreadonly,isready,isreal,issetgid,issetuid,issocket,issorted,issparse,issticky,issubnormal,issubset,issubtype,issymmetric,istaskdone,istaskstarted,istextmime,istril,istriu,isvalid,iswritable,iszero,join,joinpath,keys,keytype,kill,kron,last,lbeta,lcm,ldexp,ldltfact,ldltfact!,leading_ones,leading_zeros,length,less,lexcmp,lexless,lfact,lgamma,lgamma_fast,linearindices,linreg,linspace,listen,listenany,lock,log,log10,log10_fast,log1p,log1p_fast,log2,log2_fast,log_fast,logabsdet,logdet,logging,logm,logspace,lpad,lq,lqfact,lqfact!,lstat,lstrip,ltoh,lu,lufact,lufact!,lyap,macroexpand,map,map!,mapfoldl,mapfoldr,mapreduce,mapreducedim,mapslices,mark,match,matchall,max,max_fast,maxabs,maximum,maximum!,maxintfloat,mean,mean!,median,median!,merge,merge!,method_exists,methods,methodswith,middle,midpoints,mimewritable,min,min_fast,minabs,minimum,minimum!,minmax,minmax_fast,missing,mkdir,mkpath,mktemp,mktempdir,mod,mod1,mod2pi,modf,module_name,module_parent,mtime,muladd,mv,names,nb_available,ncodeunits,ndigits,ndims,next,nextfloat,nextind,nextpow,nextpow2,nextprod,nnz,nonzeros,norm,normalize,normalize!,normpath,notify,ntoh,ntuple,nullspace,num,num2hex,numerator,nzrange,object_id,occursin,oct,oftype,one,ones,oneunit,open,operm,ordschur,ordschur!,pairs,parent,parentindexes,parentindices,parse,partialsort,partialsort!,partialsortperm,partialsortperm!,peakflops,permute,permute!,permutedims,permutedims!,pi,pinv,pipeline,plan_bfft,plan_bfft!,plan_brfft,plan_dct,plan_dct!,plan_fft,plan_fft!,plan_idct,plan_idct!,plan_ifft,plan_ifft!,plan_irfft,plan_rfft,pointer,pointer_from_objref,pop!,popdisplay,popfirst!,position,pow_fast,powermod,precision,precompile,prepend!,prevfloat,prevind,prevpow,prevpow2,print,print_shortest,print_with_color,println,process_exited,process_running,prod,prod!,produce,promote,promote_rule,promote_shape,promote_type,push!,pushdisplay,pushfirst!,put!,pwd,qr,qrfact,qrfact!,quantile,quantile!,quit,rad2deg,rand,rand!,randcycle,randcycle!,randexp,randexp!,randjump,randn,randn!,randperm,randperm!,randstring,randsubseq,randsubseq!,range,rank,rationalize,read,read!,readandwrite,readavailable,readbytes!,readchomp,readdir,readline,readlines,readlink,readstring,readuntil,real,realmax,realmin,realpath,recv,recvfrom,redirect_stderr,redirect_stdin,redirect_stdout,redisplay,reduce,reducedim,reenable_sigint,reim,reinterpret,reload,relpath,rem2pi,repeat,replace,replace!,repmat,repr,reprmime,reset,reshape,resize!,rethrow,retry,reverse,reverse!,reverseind,rfft,rm,rol,rol!,ror,ror!,rot180,rotl90,rotr90,round,rounding,rowvals,rpad,rsearch,rsearchindex,rsplit,rstrip,run,scale!,schedule,schur,schurfact,schurfact!,search,searchindex,searchsorted,searchsortedfirst,searchsortedlast,sec,secd,sech,seek,seekend,seekstart,select,select!,selectperm,selectperm!,send,serialize,set_zero_subnormals,setdiff,setdiff!,setenv,setindex!,setprecision,setrounding,shift!,show,showall,showcompact,showerror,shuffle,shuffle!,sign,signbit,signed,signif,significand,similar,sin,sin_fast,sinc,sincos,sind,sinh,sinh_fast,sinpi,size,sizehint!,sizeof,skip,skipchars,skipmissing,sleep,slicedim,sort,sort!,sortcols,sortperm,sortperm!,sortrows,sparse,sparsevec,spawn,spdiagm,speye,splice!,split,splitdir,splitdrive,splitext,spones,sprand,sprandn,sprint,spzeros,sqrt,sqrt_fast,sqrtm,squeeze,srand,stacktrace,start,startswith,stat,std,stdm,step,stride,strides,string,stringmime,strip,strwidth,sub2ind,subtypes,success,sum,sum!,sumabs,sumabs2,summary,super,supertype,svd,svdfact,svdfact!,svdvals,svdvals!,sylvester,symdiff,symdiff!,symlink,systemerror,take!,takebuf_array,takebuf_string,tan,tan_fast,tand,tanh,tanh_fast,task_local_storage,tempdir,tempname,thisind,tic,time,time_ns,timedwait,to_indices,toc,toq,touch,trace,trailing_ones,trailing_zeros,transcode,transpose,transpose!,tril,tril!,triu,triu!,trues,trunc,truncate,trylock,tryparse,typeintersect,typejoin,typemax,typemin,unescape_string,union,union!,unique,unique!,unlock,unmark,unsafe_copy!,unsafe_copyto!,unsafe_load,unsafe_pointer_to_objref,unsafe_read,unsafe_store!,unsafe_string,unsafe_trunc,unsafe_wrap,unsafe_write,unshift!,unsigned,uperm,valtype,values,var,varinfo,varm,vcat,vec,vecdot,vecnorm,versioninfo,view,wait,walkdir,warn,which,whos,widemul,widen,withenv,workspace,write,xcorr,xor,yield,yieldto,zero,zeros,zip,broadcast_getindex,broadcast_indices,broadcast_setindex!,broadcast_similar,dotview,apropos,doc,countfrom,cycle,drop,enumerate,flatten,partition,product,repeated,rest,take,zip,get_creds!,with,calloc,errno,flush_cstdio,free,gethostname,getpid,malloc,realloc,strerror,strftime,strptime,systemsleep,time,transcode,dlclose,dlext,dllist,dlopen,dlopen_e,dlpath,dlsym,dlsym_e,find_library,adjoint,adjoint!,axpby!,axpy!,bkfact,bkfact!,chol,cholfact,cholfact!,cond,condskeel,copy_transpose!,copyto!,cross,det,diag,diagind,diagm,diff,dot,eig,eigfact,eigfact!,eigmax,eigmin,eigvals,eigvals!,eigvecs,factorize,getq,givens,gradient,hessfact,hessfact!,isdiag,ishermitian,isposdef,isposdef!,issuccess,issymmetric,istril,istriu,kron,ldltfact,ldltfact!,linreg,logabsdet,logdet,lq,lqfact,lqfact!,lu,lufact,lufact!,lyap,norm,normalize,normalize!,nullspace,ordschur,ordschur!,peakflops,pinv,qr,qrfact,qrfact!,rank,scale!,schur,schurfact,schurfact!,svd,svdfact,svdfact!,svdvals,svdvals!,sylvester,trace,transpose,transpose!,transpose_type,tril,tril!,triu,triu!,vecdot,vecnorm,html,latex,license,readme,isexpr,quot,show_sexpr,add,available,build,checkout,clone,dir,free,init,installed,pin,resolve,rm,setprotocol!,status,test,update,deserialize,serialize,blkdiag,droptol!,dropzeros,dropzeros!,issparse,nnz,nonzeros,nzrange,permute,rowvals,sparse,sparsevec,spdiagm,spones,sprand,sprandn,spzeros,catch_stacktrace,stacktrace,cpu_info,cpu_summary,free_memory,isapple,isbsd,islinux,isunix,iswindows,loadavg,total_memory,uptime,atomic_add!,atomic_and!,atomic_cas!,atomic_fence,atomic_max!,atomic_min!,atomic_nand!,atomic_or!,atomic_sub!,atomic_xchg!,atomic_xor!,nthreads,threadid,applicable,eval,fieldtype,getfield,invoke,isa,isdefined,nfields,nothing,setfield!,throw,tuple,typeassert,typeof,uninitialized,undef},%
    keywords=[2]{AbstractArray,AbstractChannel,AbstractDict,AbstractDisplay,AbstractFloat,AbstractIrrational,AbstractMatrix,AbstractRNG,AbstractRange,AbstractSerializer,AbstractSet,AbstractSparseArray,AbstractSparseMatrix,AbstractSparseVector,AbstractString,AbstractUnitRange,AbstractVecOrMat,AbstractVector,Adjoint,Any,ArgumentError,Array,AssertionError,Bidiagonal,BigFloat,BigInt,BitArray,BitMatrix,BitSet,BitVector,Bool,BoundsError,BufferStream,CapturedException,CartesianIndex,CartesianIndices,Cchar,Cdouble,Cfloat,Channel,Char,Cint,Cintmax_t,Clong,Clonglong,Cmd,CodeInfo,Colon,Complex,ComplexF16,ComplexF32,ComplexF64,CompositeException,Condition,ConjArray,ConjMatrix,ConjVector,Cptrdiff_t,Cshort,Csize_t,Cssize_t,Cstring,Cuchar,Cuint,Cuintmax_t,Culong,Culonglong,Cushort,Cvoid,Cwchar_t,Cwstring,DataType,DenseArray,DenseMatrix,DenseVecOrMat,DenseVector,Diagonal,Dict,DimensionMismatch,Dims,DivideError,DomainError,EOFError,EachLine,Enum,Enumerate,ErrorException,Exception,ExponentialBackOff,Expr,Factorization,Float16,Float32,Float64,Function,GlobalRef,GotoNode,HTML,Hermitian,IO,IOBuffer,IOContext,IOStream,IPAddr,IPv4,IPv6,IndexCartesian,IndexLinear,IndexStyle,InexactError,InitError,Int,Int128,Int16,Int32,Int64,Int8,Integer,InterruptException,InvalidStateException,Irrational,KeyError,LabelNode,LinSpace,LineNumberNode,LinearIndices,LoadError,LowerTriangular,MIME,Matrix,MersenneTwister,Method,MethodError,MethodTable,Missing,MissingException,Module,NTuple,NamedTuple,NewvarNode,Nothing,Number,ObjectIdDict,OrdinalRange,OutOfMemoryError,OverflowError,Pair,PartialQuickSort,PermutedDimsArray,Pipe,Ptr,QuoteNode,RandomDevice,Rational,RawFD,ReadOnlyMemoryError,Real,ReentrantLock,Ref,Regex,RegexMatch,RoundingMode,RowVector,SSAValue,SegmentationFault,SerializationState,Set,Signed,SimpleVector,Slot,SlotNumber,Some,SparseMatrixCSC,SparseVector,StackFrame,StackOverflowError,StackTrace,StepRange,StepRangeLen,StridedArray,StridedMatrix,StridedVecOrMat,StridedVector,String,StringIndexError,SubArray,SubString,SymTridiagonal,Symbol,Symmetric,SystemError,TCPSocket,Task,Text,TextDisplay,Timer,Transpose,Tridiagonal,Tuple,Type,TypeError,TypeMapEntry,TypeMapLevel,TypeName,TypeVar,TypedSlot,UDPSocket,UInt,UInt128,UInt16,UInt32,UInt64,UInt8,UndefRefError,UndefVarError,UniformScaling,Uninitialized,Union,UnionAll,UnitRange,Unsigned,UpperTriangular,Val,Vararg,VecElement,VecOrMat,Vector,VersionNumber,WeakKeyDict,WeakRef,BLAS,Base,Broadcast,DFT,Docs,Iterators,LAPACK,LibGit2,Libc,Libdl,LinAlg,Markdown,Meta,Operators,Pkg,Serializer,SparseArrays,StackTraces,Sys,Threads,Core,Main},%
    keywords=[1]{true,false,nothing,missing,im,uninitialized,NaN,NaN16,NaN32,NaN64,Inf,Inf16,Inf32,Inf64,ARGS,C_NULL,ENDIAN_BOM,ENV,LOAD_PATH,PROGRAM_FILE,STDERR,STDIN,STDOUT,VERSION},
    keywords=[1]{mutable,immutable,struct,begin,end,function,macro,quote,let,local,global,const,abstract,module,baremodule,using,import,export,in,if,else,elseif,for,while,do,try,type,catch,finally,return,break,continue},%
    sensitive=true,
    morecomment=[l]{\#},
    morecomment=[n]{\#=}{=\#},
    morestring=[s]{"}{"},
    morestring=[m]{'}{'},
    literate=*{-}{-}1,
    alsoletter=!?
}

\lstdefinestyle{julia}{
    basicstyle       = \ttfamily\small\color[HTML]{333333},
    numberstyle      = \ttfamily\scriptsize\color[HTML]{7F7F7F},
    keywordstyle     = [1]{\ttfamily\color[HTML]{2D2F92}},
    keywordstyle     = [2]{\color[HTML]{006795}},
    keywordstyle     = [3]{\color[HTML]{9D5C02}},
    stringstyle      = \ttfamily\color[HTML]{880000},
    commentstyle     = \color[rgb]{0,0.6,0},
    rulecolor        = \color[HTML]{000000},
    frame=lines,
    xleftmargin=5pt,
    framexleftmargin=5pt,
    framextopmargin=2pt,
    framexbottommargin=2pt,
    tabsize=4,
    captionpos=b,
    breaklines=true,
    breakatwhitespace=false,
    showstringspaces=false,
    showspaces=false,
    showtabs=false,
    columns=fullflexible,
    keepspaces=true,
    numbers=none
}

\lstdefinelanguage{JuliaLocal}{
    language = Julia, % inherit Julia lang. to add keywords
    morekeywords = [3]{ominmax, satisfaction_probability, control_synthesis, value_iteration, read_prism_file, read_bmdp_tool_file, read_intervalmdp_jl_file, write_intervalmdp_jl_file, cumulative_sum, is_functional, cu, omaximization, write_bmdp_tool, write_prism_file, write_bmdp_tool_file, read_intervalmdp_jl_model, read_intervalmdp_jl_spec, read_intervalmdp_jl, write_intervalmdp_jl_model, write_intervalmdp_jl_spec}, % define more functions
    morekeywords = [2]{IntervalProbabilities, IntervalMarkovDecisionProcess, IntervalMarkovChain, IntervalMDP, Data, CUDA, FiniteTimeReachability, InfiniteTimeReachability, Problem, Specification, FiniteTimeReward}, % define more types and modules
}

\definecolor{delim}{RGB}{20,105,176}
\definecolor{numb}{RGB}{106, 109, 32}
\definecolor{string}{rgb}{0.64,0.08,0.08}

\lstdefinelanguage{json}{
    numbers=none,
    frame=none,
    showspaces=false,
    showtabs=false,
    breaklines=true,
    postbreak=\raisebox{0ex}[0ex][0ex]{\ensuremath{\color{gray}\hookrightarrow\space}},
    breakatwhitespace=true,
    basicstyle=\ttfamily\small,
    upquote=true,
    morestring=[b]",
    stringstyle=\color{string},
    literate=
     *{0}{{{\color{numb}0}}}{1}
      {1}{{{\color{numb}1}}}{1}
      {2}{{{\color{numb}2}}}{1}
      {3}{{{\color{numb}3}}}{1}
      {4}{{{\color{numb}4}}}{1}
      {5}{{{\color{numb}5}}}{1}
      {6}{{{\color{numb}6}}}{1}
      {7}{{{\color{numb}7}}}{1}
      {8}{{{\color{numb}8}}}{1}
      {9}{{{\color{numb}9}}}{1}
      {\{}{{{\color{delim}{\{}}}}{1}
      {\}}{{{\color{delim}{\}}}}}{1}
      {[}{{{\color{delim}{[}}}}{1}
      {]}{{{\color{delim}{]}}}}{1},
}

\usepackage{xspace}
\usepackage[acronym]{glossaries}

\usepackage{etoolbox}
\newbool{inproceedings}
\boolfalse{inproceedings}

\input{macros.tex}
\input{acronyms.tex}

\begin{document}
\begin{frontmatter}

\title{\IMDPjl: Accelerated Value Iteration for Interval Markov Decision Processes} 

% \thanks[footnoteinfo]{Sponsor and financial support acknowledgment
% goes here. Paper titles should be written in uppercase and lowercase
% letters, not all uppercase.}

\author[First]{Frederik Baymler Mathiesen} 
\author[Second]{Morteza Lahijanian} 
\author[Third]{Luca Laurenti}

\address[First]{Delft University of Technology, the Netherlands (e-mail: f.b.mathiesen@tudelft.nl)}
\address[Second]{University of Colorado, Boulder, USA (e-mail: morteza.lahijanian@colorado.edu)}
\address[Third]{Delft University of Technology, the Netherlands (e-mail: l.laurenti@tudelft.nl)}

\begin{abstract}                % Abstract of not more than 250 words.
In this paper, we present \IMDPjl, a Julia package for probabilistic analysis of interval Markov Decision Processes (IMDPs). \IMDPjl facilitates the synthesis of optimal strategies and verification of IMDPs against reachability specifications and discounted reward properties. The library supports sparse matrices and is compatible with data formats from common tools for the analysis of probabilistic models, such as PRISM. A key feature of \IMDPjl is that it presents both a multi-threaded CPU and a GPU-accelerated implementation of value iteration algorithms for IMDPs. In particular,  \IMDPjl takes advantage of the Julia type system and the inherently parallelizable nature of value iteration to improve the efficiency of performing analysis of IMDPs. On a set of examples, we show that \IMDPjl substantially outperforms existing tools for verification and strategy synthesis for IMDPs in both computation time and memory consumption.
\end{abstract}

\begin{keyword}
Markov Decision Processes, Robust Value Iteration, Reachability, Control Synthesis, Verification.
\end{keyword}

\end{frontmatter}
%===============================================================================

\section{Introduction}
\Glspl{imdp} \citep{givan2000bounded} are a class of uncertain \glspl{mdp}, where the transition probabilities between each pair of states are uncertain and lie in some probability interval. Due to their modeling flexibility and the existence of efficient robust value iteration algorithms \citep{givan2000bounded, HADDAD2018111}, \glspl{imdp} have recently gained increasing attention both in the computer science and control communities \citep{HADDAD2018111, cauchi2019efficiency,laurenti2023unifying, lavaei2022automated}. In particular, \glspl{imdp} are popular for abstraction-based verification and control of stochastic hybrid systems \cite{cauchi2019efficiency, lavaei2022automated}. However, existing tools for performing verification and strategy synthesis of \glspl{imdp} lack parallelization via multi-threading and GPU-acceleration, under-utilizing available hardware.

%Model checking for this class of models, similarly to regular \glspl{mdp}, reduces to value iteration, although with the addition of an adversary to select probabilities within the uncertainty set. One important type of model checking is probabilistic reachability, specifically as it enables verification of more complex specifications via a product construction.
%State-of-the-art probabilistic model checkers for \glspl{imdp} are PRISM (\cite{kwiatkowska2011prism}) and bmdp-tool (\cite{morteza_lahijanian_bmdp_tool}). Neither of these tools parallellize the verification however, leaving a gap for our tool \IMDPjl.

In this paper, we introduce \IMDPjl, which is a Julia package for reachability analysis of \glspl{imdp}. The tool supports both reachability specifications and discounted-reward properties, and allows the user to query for both optimal strategies and quantitative values of satisfaction. The package contains both a CPU implementation and an implementation that uses \glspl{gpu} allowing one to use CUDA-capable hardware \citep{cuda} to accelerate the computation. The package is developed in Julia, which is a modern programming language that targets the scientific community \citep{bezanson2012julia}. It enables both fast prototyping and accelerated code to be written in the same language, including the ability to write CUDA kernels for custom accelerated computations. %The latter is necessary to achieve the speed gains of \IMDPjl. 
Furthermore, through Julia's parametric typing, \IMDPjl supports single- and double-precision floating point numbers, as well as rational numbers for exact arithmetic with transition probabilities.

We evaluate \IMDPjl on various benchmarks and compare it with PRISM \citep{kwiatkowska2011prism} and bdmp-tool \citep{morteza_lahijanian_bmdp_tool}, the only tools available to perform reachability analysis for \IMDPjl to the best of our knowledge. The benchmarks include 35 \glspl{imdp} taken from the literature, with the total number of transitions between states ranging from a few tens for the smaller models to tens of millions for the larger models. The empirical analysis shows that \IMDPjl CPU implementation is on average $2$-$4\times$ faster compared to the state of the art, while the \gls{gpu} implementation can achieve speed-ups of various orders of magnitude on the larger systems. Furthermore, because of the use of sparse matrices and the Julia type system, in all cases, \IMDPjl requires less memory compared to PRISM and bmdp-tool.

The paper is organized as follows.  First, in Section \ref{sec:IMDPOverview}, we formally introduce \glspl{imdp} and robust value iteration for \glspl{imdp}. Then, in Section \ref{sec:overview}, we give an overview of \IMDPjl and describe how \glspl{imdp} can be created and stored in the tool and how to perform strategy synthesis and verification. In Section \ref{sec:valueIteration}, we detail our algorithmic approach to robust value iteration on \glspl{gpu}. Finally, in Section \ref{sec:experiments}, we illustrate the effectiveness of \IMDPjl on various benchmarks.

\section{The goal of \IMDPjl: Robust Value Iteration for IMDPs}
\label{sec:IMDPOverview}
\Acrfullpl{imdp}, also called bounded-parameter \glspl{mdp} \citep{givan2000bounded}, are a generalization of \glspl{mdp}, where the transition probabilities between each pair of states are not known exactly, but they are constrained to be in some independent probability intervals. 
Formally, an \gls{imdp} $M$ is a tuple $M = (S, A, \xoverline{P}, \underline{P})$, where
% \LL{Do we need the initial state in here? Why do not we simply return the probability for each state? I know this is how they do it in Prism, but then we should specify that we also have a command to compute the property for all initial states at once}
\begin{itemize}
    \item $S$ is a finite set of states,
    % \item $s_0 \in S$ is the initial state,
    \item $A$ is a finite set of actions, where the set of actions available at state $s \in S$ is denoted by $\mathcal{A}(s) \subseteq A$,
    \item $\underline{P}: S \times A \times S \to [0,1]$ is a function, where $\underline{P}(s,a,s')$ defines the lower bound of the transition probability from state $s\in S$ to state $s'\in S$ under action $a \in \mathcal{A}(s)$,
    \item $\xoverline{P}: S \times A \times S \to [0,1]$ is a function, where $\xoverline{P}(s,a,s')$ defines the upper bound of the transition probability from state $s\in S$ to state $s'\in S$ under action $a \in \mathcal{A}(s)$.
\end{itemize}
For each state-action pair $(s,a) \in S \times A$ where $a \in \mathcal{A}(s)$, it holds that $\sum_{s'\in S} \underline{P}(s,a,s') \leq 1 \leq \sum_{s'\in S} \xoverline{P}(s,a,s')$ and a transition probability distribution $p_{s,a}:S\to[0,1]$ is called \emph{feasible} if $\underline{P}(s,a,s')\leq p_{s,a}(s')\leq\xoverline{P}(s,a,s')$ for all $s'\in S$. The set of all feasible distributions for the state-action pair $(s,a)$ is denoted by $\Gamma_{s,a}$.
% We define $\Gamma = \{\Gamma_{q,a}:(q,a)\inS\timesA\}$ to be the set of all feasible distributions for all state-action pairs.

A \emph{path} of an \gls{imdp} is a sequence of states and actions $\omega = (s_0,a_0),(s_1,a_1),\dots$, where $(s_i,a_i)\in S \times A$ and $a_i \in \mathcal{A}(s_i)$. We denote by $\omega(k) = s_k$ the state of the path at time $k \in \mathbb{N}_0$ and by $\Omega$ the set of all paths.  
A \emph{strategy} or \emph{policy} for an \gls{imdp} is a function $\pi$ that assigns an action to a given state of an \gls{imdp}. In \IMDPjl, we focus on \emph{time-dependent} 
strategies, i.e., $\pi: S\times \mathbb{N}_0 \to A$, as they suffice for optimality for reachability and discounted cumulative reward properties \citep{givan2000bounded,delimpaltadakis2023interval}. In the special case that $\pi$ does not depend on time and solely depends on the state, it is called a \emph{stationary} strategy, i.e., $\pi(s,k)=\pi(s,k')$ for all $k,k' \in \mathbb{N}_0$ and all $s \in S$. Similar to a strategy, an adversary $\eta$ is a function that assigns a feasible distribution to a given state \citep{givan2000bounded}. Given a strategy and an adversary, an \gls{imdp} collapses to a finite Markov chain.

In \IMDPjl, we consider both reachability and discounted-reward properties. In the rest of this section, we focus on the reachability problem; the discounted reward case follows similarly. For reachability on \glspl{imdp}, we define a goal region $G\subseteq S$ and a time horizon $K\in \mathbb{N}_0 \cup \{\infty\}$,
% \footnote{In the current version of \IMDPjl for reachability properties, as common in other tools, for $K=\infty$, we require that any transition with a non-zero upper bound also has a non-zero lower bound. The more general case requires to identify maximal end components of the \gls{imdp} and is the subject of future work.
% \ml{I believe we do not need this footnote. For reachability, we don't need to analyze end components.  That's only needed for satisfying Buchi/Rabin conditions (LTL properties), which is beyond reachability.}
% }
then the objective is to solve the following optimization problems:
% \begin{align}
% \label{eq:PessimisticReward}
% & \max_{\pi}\min_{\eta}\mathbb{P}_{\pi,\eta }\left[\omega \in \Omega \mid \exists k \in [0,K], \, \omega(k)\in G \right], \\
% &\max_{\pi}\max_{\eta}\mathbb{P}_{\pi,\eta }\left[\omega \in \Omega \mid \exists k \in [0,K], \, \omega(k)\in G  \right],
% \label{eq:optimisitnReward}
% \end{align}
% where $\mathbb{P}_{\pi,\eta }$ is the probability of the  Markov chain induced by strategy $\pi$ and adversary $\eta$.
% Eqn. \eqref{eq:PessimisticReward} is called  \emph{pessimistic} optimal probability (or reward) and Eqn. \eqref{eq:PessimisticReward} is called \emph{optimistic} optimal probability (or reward).
% Computing Eqn. \eqref{eq:PessimisticReward} can be done by solving the following robust value iteration (similar value iteration can be found for the optimistic case, by replacing the $\min$ with the $\max$)
% \begin{align}
% \nonumber
%  V_{0}(s)&=\mathbf{1}_{G}(s) \\
% \label{Eq:RobustValueITeration}
%     V_{k}(s) &= \mathbf{1}_{G}(s) +  \mathbf{1}_{S\setminus G}(s)\max_{a \in A}\min_{p_{s,a}\in \Gamma_{s,a}} \sum_{s' \in S} V_{k-1}(s') p_{s,a}(s'),
% \end{align}
% where $\mathbf{1}_{G}(s)\begin{cases} 1 \quad& \text{if } s \in G \\ 0 \quad & \text{otherwise}
% \end{cases}$ is the indicator function for set $G$. 
\begin{align}
\label{eq:PessimisticReward}
    \opta_{\pi}\opte_{\eta} \mathbb{P}_{\pi,\eta }\left[\omega \in \Omega \mid \exists k \in [0,K], \, \omega(k)\in G \right],
    % &\max_{\pi}\max_{\eta}\mathbb{P}_{\pi,\eta }\left[\omega \in \Omega \mid \exists k \in [0,K], \, \omega(k)\in G  \right],
    % \label{eq:optimisitnReward}
\end{align}
where $\opta,\opte \in \{\min, \max\}$ and $\mathbb{P}_{\pi,\eta }$ is the probability of the Markov chain induced by strategy $\pi$ and adversary $\eta$.
When $\opte = \min$, Eqn.~\eqref{eq:PessimisticReward} is called optimal \emph{pessimistic} probability (or reward), and conversely Eqn.~\eqref{eq:PessimisticReward} is called optimal \emph{optimistic} probability (or reward) when $\opte = \max$.
The choice of the min/max for the action and pessimistic/optimistic probability depends on the application. 
For instance, in robust strategy synthesis, the interest is often in maximizing pessimistic probability, i.e., $\opta = \max$ and $\opte = \min$.

Eqn. \eqref{eq:PessimisticReward} can be computed by solving the following value iteration:
\begin{equation}\label{Eq:RobustValueITeration}
    \begin{aligned}
        V_{0}(s) &=\mathbf{1}_{G}(s) \\
        V_{k}(s) &= \mathbf{1}_{G}(s) \;+ \\
                 &\phantom{=}\;\; \mathbf{1}_{S\setminus G}(s)\opta_{a \in \mathcal{A}(s)}\opte_{p_{s,a}\in \Gamma_{s,a}} \sum_{s' \in S} V_{k-1}(s') p_{s,a}(s'),
    \end{aligned}
\end{equation}
where $\mathbf{1}_{G}(s) = \begin{cases} 1 \quad& \text{if } s \in G \\ 0 \quad & \text{otherwise}
\end{cases}$ is the indicator function for set $G$. 
% \FM{Should we highlight that this is propagation backwards through time despite indexing in increasing order?}\LL{I formulated it in this way as it is more common in some literature, especially when you look at convergence for infinite-horizon properties and consequently you do not have a final time from which you would start to go back. In any case, to clarify, I think it is enough to say that this is a dynamic programming problem and refer to a good reference.  }
\IMDPjl solves Eqn. \eqref{Eq:RobustValueITeration} iteratively over time via linear programming. In Section \ref{sec:valueIteration}, we show how its solution can be efficiently parallelized to take advantage of \gls{gpu} hardware architectures.

%\begin{remark}
%Remark how it extends to reachability etc etc
%\end{remark}

%Given a strategy, an IMDP defines a family of Markov chains. An adversary is a function  $\eta: \Omega \rightarrow \Gamma$ that for any path assign a feasible distribution. Given a strategy and an adversay an IMDP collpases to a finite Markov chain. 

%%Verification of \glspl{imdp} using value iteration works by specifying a (time-varying) reward function $R : S \times \mathbb{N}_0 \to \mathbb{R}$ and a discount factor $\gamma$ according to the property desired. Probabilistic reachability can be describe by the reward function $R(s, H) = \mathbf{1}_T(s)$ where $T$ is the target set and $H$ is the (possibly infinite) time horizon, and $R(s, k) = 0$ for $k < H$, and $\gamma = 1$. Therefore, we may generally describe the problem in terms of a reward function. 

%Let an \gls{imdp} $M$, a reward function $R$, and a discount factor $\gamma$ be given. Then the problem is to compute a policy $\pi$ that minimizes or maximizes the pessimistic or optimistic expected reward
%\begin{equation}
%    V_0^\pi(s) = \mathbb{E}_{\omega \sim \mathbb{P}^\pi(\cdot)}\left[\sum_{k = 0}^\infty \gamma^k R(\sigma_k, k) \right]
%\end{equation}

\section{Overview of \IMDPjl}
\label{sec:overview}
%The goal of \IMDPjl is verification of reachability specifications and control synthesis to optimize reachability probabilities in systems modeled as \glspl{imdp}.
\IMDPjl is a Julia package that introduces parallelization and \gls{gpu}-powered processing to perform value iteration 
% (i.e., to solve Eqn. \eqref{Eq:RobustValueITeration}) 
in Eqn.~\eqref{Eq:RobustValueITeration}, offering efficiency for verification and strategy synthesis for \glspl{imdp}.  
\IMDPjl has the following main features:
\begin{itemize}
    \item value iteration and strategy synthesis for all combinations of $\opta$ and $\opte$,
    \item dense and sparse matrix support,
    \item customizable numerical precision,
    \item multi-threaded CPU and CUDA-accelerated value iteration, and
    \item data loading and writing in various formats (e.g., PRISM, bmdp-tool, and \IMDPjl).
\end{itemize}

In this section, we show how to create an \gls{imdp} model in \IMDPjl
% . Then, in Section \ref{sec:spec_and_verirification}, we focus on how to 
and use \IMDPjl for verification and synthesis.
The source code can be found at \url{https://github.com/Zinoex/IntervalMDP.jl}.

% \subsection{Installation}
% \label{sec:installation}
% \IMDPjl is a Julia package. Consequently, to install and import the package, it is enough to run the following commands in Julia:
% \begin{lstlisting}[language=JuliaLocal, style=julia]
% using Pkg
% Pkg.add("IntervalMDP")

% using IntervalMDP, IntervalMDP.Data
% \end{lstlisting}
% The module \lstinline[language=JuliaLocal, style=julia]|IntervalMDP| contains structures and functions for modeling, value iteration, and strategy synthesis. The submodule \lstinline[language=JuliaLocal, style=julia]|IntervalMDP.Data| contains functions relevant to reading and writing \glspl{imdp} in various data formats.

\subsection{System modeling}\label{sec:system_modelling}
%Traditionally, probabilistic model checking has been developed as a standalone program, necessitating that system description including transition probabilities are written to a file by the calling software, only to be read by the model checker. However, since 
% \IMDPjl is a Julia package. Consequently, 
To create an \gls{imdp}, we can programmatically construct an 
\lstinline[language=JuliaLocal, style=julia]|IntervalMarkovDecisionProcess| object
% \footnote{The package also includes an \lstinline[language=JuliaLocal, style=julia]|IntervalMarkovChain| type to represent an \gls{imdp} with a single action. This type is slightly more efficient through better vectorization.}
and pass it to \IMDPjl.
Here, we include an example of how to construct a 3-state \gls{imdp} with the third state being a sink state. 
\begin{lstlisting}[language=JuliaLocal, style=julia]
prob1 = IntervalProbabilities(;
    lower = [0.0 0.5; 0.1 0.3; 0.2 0.1],
    upper = [0.5 0.7; 0.6 0.5; 0.7 0.3],
)
prob2 = IntervalProbabilities(;
    lower = [0.1 0.2; 0.2 0.3; 0.3 0.4],
    upper = [0.6 0.6; 0.5 0.5; 0.4 0.4],
)
prob3 = IntervalProbabilities(;
    lower = [0.0; 0.0; 1.0],
    upper = [0.0; 0.0; 1.0]
)
transition_probs = [
    ["a1", "a2"] => prob1,
    ["a1", "a2"] => prob2,
    ["sink"] => prob3
]
imdp = IntervalMarkovDecisionProcess(transition_probs)
\end{lstlisting}

\begin{figure}
    \centering
    \includegraphics[width=0.8\linewidth]{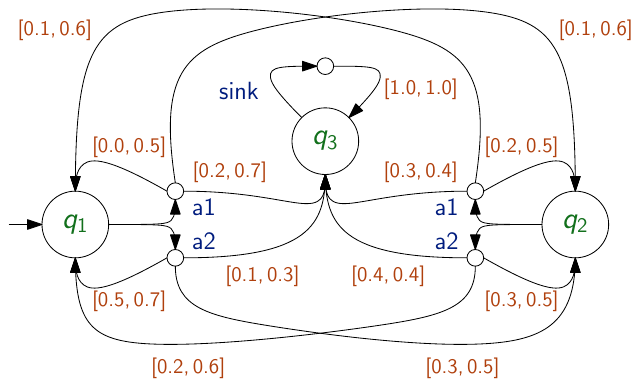}
    \caption{An example of a 3-state \gls{imdp} where state $q_3$ is a sink state.}
    \label{fig:imdp_example}
\end{figure}

Figure \ref{fig:imdp_example} shows a pictorial representation of the \gls{imdp} constructed with the code above.
The interval transition probabilities are specified as a list \lstinline[language=JuliaLocal, style=julia]|actions => transition_probs|where the upper and lower bounds are specified using dense matrices with columns representing the actions and rows the resulting state. \IMDPjl also supports sparse matrices encoded in the \gls{csc} format \citep{10.1145/1583991.1584053} for better memory usage and computational efficiency. 
%Sparse matrices in \gls{csc} format are implemented in the Julia standard library.
Note that while in the example above the actions are given as strings, namely \lstinline[language=JuliaLocal, style=julia]|"a1"|, \lstinline[language=JuliaLocal, style=julia]|"a2"|, and \lstinline[language=JuliaLocal, style=julia]|"sink"|, the package supports any arbitrary type as actions, e.g., integers and floating point numbers.

To improve compatibility with other tools,  \IMDPjl supports reading and writing \glspl{imdp} stored in various formats. 
Currently, there are three supported standards for storing \glspl{imdp}: PRISM, bmdp-tool, and \IMDPjl data formats.
The following subsections describe how to read and write in the respective formats. 
% We provide a detailed description of each format in Appendix~\ref{sec:appendix_storage_formats}.

\subsubsection{PRISM}
\IMDPjl supports reading and writing PRISM \citep{kwiatkowska2011prism} explicit data format. 
% \LL{Do we read both explicit and implicit formats? }\FM{We only support the explicit format. For IMDPs, PRISM uses the explicit engine regardless, so there are no performance differences, and reading the implicit formats is massively more complicated.}\LL{Then, I would not even talk about the discussion about implicit and explicit formats. Let's just say that we import IMDPs written in the PRISM explicit format. }
% The explicit description is more efficient to read for the class of systems described in this paper, namely because it bypasses the model parser and constructor.
% The explicit format uses a combination of variables to identify states in the \gls{imdp}.
The data format is split into 4 different files, one for the states '.sta', one for the labels '.lab', one for the transition probabilities '.tra', and one for the specification '.pctl'. Therefore, our interface for reading PRISM files takes the path without file ending and adds the appropriate ending to each of the four files.
% We support however only single variable state identifiers, as we use integer indices to identify state in \IMDPjl.
\begin{lstlisting}[language=JuliaLocal, style=julia]
# Read
problem = read_prism_file(path_without_file_ending)

# Write
write_prism_file(path_without_file_ending, problem)
\end{lstlisting}
The problem structure contains both the \gls{imdp} and the specification including whether to synthesize a maximizing or minimizing strategy and whether to use an optimistic or pessimistic adversary.
%This will be described in greater detail in Subsection \ref{sec:spec_and_verirification}

\subsubsection{bmdp-tool}
bmdp-tool \citep{morteza_lahijanian_bmdp_tool} data format is similar to the PRISM explicit format transition probability files, where transition probabilities are stored line-by-line with source, action, destination, and probability bounds in ASCII. Key differences include no explicit listing of states and the fact that it only supports reachability properties. This format lacks information about whether the reachability is finite or infinite time, and hence the reader only returns the set of terminal states.
\begin{lstlisting}[language=JuliaLocal, style=julia]
# Read
imdp, terminal_states = read_bmdp_tool_file(path)

# Write
write_bmdp_tool_file(path, problem)
\end{lstlisting}

\subsubsection{\IMDPjl}
A disadvantage of both PRISM and bmdp-tool data formats is that they store the data in ASCII where each character requires one byte of storage space. As a result, the data format requires more storage space than necessary for floating point data. \IMDPjl supports NetCDF \citep{rew2006netcdf} to store transition probabilities efficiently. NetCDF is a multi-dimensional array-centric format with readers and writers in many languages and includes lossless compression methods for even more efficient data storage.
We use JSON to store the specification, as storage space for the specification is much less a concern, and because JSON is a widely used, human-readable, file format.
\begin{lstlisting}[language=JuliaLocal, style=julia]
# Read
imdp = read_intervalmdp_jl_model(model_path)
spec = read_intervalmdp_jl_spec(spec_path)
problem = Problem(imdp, spec)

problem = read_intervalmdp_jl(model_path, spec_path)

# Write
write_intervalmdp_jl_model(model_path, imdp_or_problem)
write_intervalmdp_jl_spec(spec_path, spec_or_problem)
\end{lstlisting}
Without compression, this data format requires 50\% less space compared to both ASCII formats.

\subsection{Specification and verification}\label{sec:spec_and_verirification}
The tool minimizes or maximizes the optimistic or pessimistic probability of reaching a given set of states, or optimizes a reward.
Continuing on the example from Section \ref{sec:system_modelling}, we show how to compute the maximum pessimistic probability of reaching state 3 within 100 time steps.
\begin{lstlisting}[language=JuliaLocal, style=julia]
prop = FiniteTimeReachability([3], 100)
spec = Specification(prop, Pessimistic, Maximize)
problem = Problem(imdp, spec)

# Compute the value function
V, k, residual = value_iteration(problem)
\end{lstlisting}
Function \lstinline[language=JuliaLocal, style=julia]|value_iteration| returns a 3-tuple containing the following: the robust optimal probability (in the above example, maximum pessimistic probability) of satisfying the reachability property, the number of iterations performed (100 in the above example), and finally, the Bellman residual for the last iteration. That is, the state-wise difference between the value function for the last and the second to last iteration.
For a finite horizon property, the number of iterations is fixed, while for an infinite horizon property, a user may specify a tolerance for convergence, measured by the maximum Bellman residual.
% Underlying it is calling \lstinline[language=JuliaLocal, style=julia]|value_iteration|, which is also exported as an interface. 
% \FM{I have modified the text above because what was written before made no sense syntactically or semantically, and was factually incorrect to what the function returns.}

Instead, to compute robust optimal discounted cumulative reward property, \IMDPjl can be used as follows:
\begin{lstlisting}[language=JuliaLocal, style=julia]
rewards = [1.0, 2.0, 3.0]
discount = 0.95
prop = FiniteTimeReward(rewards, discount, 100)
\end{lstlisting}
The length of the list specifying the rewards must match the number of states in the \gls{imdp} and the discount must be between $0$ and $1$. Then the specification and problem construction follow as with a reachability property.

Using the library, we may also synthesize the optimal policy. 
\begin{lstlisting}[language=JuliaLocal, style=julia]
# Compute optimal finite-time policy
prop = FiniteTimeReachability([3], 100)
spec = Specification(prop, Pessimistic, Maximize)
problem = Problem(imdp, spec)
time_dependent_policy = control_synthesis(problem)

# Compute optimal infinite-time policy
prop = InfiniteTimeReachability([3], 1e-6)
spec = Specification(prop, Pessimistic, Maximize)
problem = Problem(imdp, spec)
stationary_policy = control_synthesis(problem)
\end{lstlisting}

% \LL{Next Sentence is not very clear. I would rephrase by saying say that in the first case the policy will be time dependent and thus \IMDPjl will return a $A^{|S| \times \mathbb{K}}$ matrix, while in the second case the optimal policy is  stationary}

In the first case, with a finite time horizon property, the optimal strategy is time dependent, and thus \IMDPjl returns a matrix of size ${|S| \times K}$ where $K$ is the time horizon,  whose entry in the $i$-th row and $j$-th column is the action for state $i$ at time step $j$.
In the second case, the optimal strategy is stationary and \IMDPjl returns a vector of length $|S|$, where the $i$-th entry is the action for state $i$.
% A time-varying policy is a matrix $A^{|S| \times \mathbb{K}}$ where the rows represent the states and the columns represent discrete time in increasing order.
% As previously mentioned, a stationary policy does not depend on time and it is thus a vector $A^{|S|}$.

A core feature of \IMDPjl is GPU-accelerated algorithms.
To enable the use of CUDA, we only need to transfer the problem to the \gls{gpu}. 
Note that \glspl{gpu} generally have less memory than what is available to CPUs, and thus it is highly recommended the use of the sparse format for \glspl{gpu}.

For the example above, the following code enables the use of a \gls{gpu}, if available:
\begin{lstlisting}[language=JuliaLocal, style=julia]
using CUDA

prop = FiniteTimeReachability([3], 100)
spec = Specification(prop, Pessimistic, Maximize)
problem = Problem(imdp, spec)

if CUDA.functional()
    problem = IntervalMDP.cu(problem)
end

V, k, residual = value_iteration(problem)
\end{lstlisting}
This trivial modification allows improvements in terms of computational time of $50$-$200\times$ as is illustrated in Section \ref{sec:experiments}. Note that \lstinline[language=JuliaLocal, style=julia]|IntervalMDP.cu| is opinionated to \texttt{Float64} values and \texttt{Int32} indices to balance numerical errors and performance. 
However, other options for value and index types are available via the \texttt{Adapt.jl} package.

\section{Value Iteration on CUDA-capable GPUs}
\label{sec:valueIteration}

In this section, we discuss our algorithmic approach to solve Eqn.~\eqref{eq:PessimisticReward} 
% and \eqref{eq:optimisitnReward} 
on CUDA-capable hardware. 
% For completeness, we include a short description of the CUDA programming model in Appendix~\ref{sec:appendix_cuda}. 
For the sake of simplicity of presentation, we focus only on the 
% pessimistic optimal probability case (Eqn. \eqref{eq:PessimisticReward}); 
maximum pessimistic probability ($\opta = \max$ and $\opte = \min$ in Eqn. \eqref{eq:PessimisticReward}) case; the other cases follow similarly. 
Recall from Eqn. \eqref{eq:PessimisticReward} that computing the optimal pessimistic reward reduces to iteratively solving the following problem for each $s\in S$ and for each $k$
\begin{align}
\label{eqn:StepValueIteration}
   & \max_{a \in \mathcal{A}(s)}\min_{p_{s,a}\in \Gamma_{s,a}} \sum_{s' \in S} V_{k}(s') p_{s,a}(s').
\end{align}
As Eqn. \eqref{eqn:StepValueIteration} is solved independently for each state $s$, its solution at different states can be trivially parallelized. However, in what follows we show how in \IMDPjl, the solution of robust value iteration is parallelized further.

\begin{figure}
    \centering
    \includegraphics[width=0.8\linewidth]{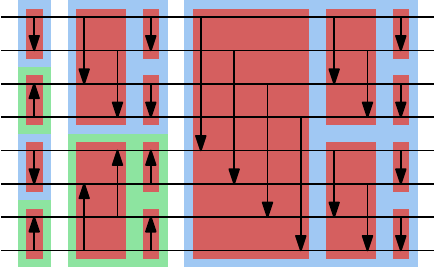}
    \caption{Bitonic sorting network structure in $\log_2(n)$ major rounds, each of which is up to $\log_2(n)$ minor rounds. Each arrow represents a comparison and  points towards the larger element. After each major round, subsets of increasing size are bitonic (increasing order, then decreasing of the same size).}
    \label{fig:bitonic_sort}
\end{figure}

The inner problem in Eqn. \eqref{eqn:StepValueIteration} is a linear problem that must be solved for every $a\in \mathcal{A}(s)$. 
We solve this problem using the O-maximization procedure introduced in \citep{givan2000bounded,7029024}. 
The O-maximization procedure is divided into two phases: ordering of states based on the value of $V_k$ and then assigning the least probability mass possible to states with high $V_k$.
Both phases can benefit from parallelization on a \gls{gpu}. For ordering the states based on the value of the value function at the current time step $V_k$, we apply existing parallel algorithms that are particularly suited to run efficiently on a \gls{gpu}. 
In particular, in \IMDPjl, we use bitonic sort \citep{10.1007/978-3-642-14390-8_42}, which is a parallel sorting algorithm with $O(\log_2(n)^2)$ latency\footnote{Note that when assessing parallel algorithms, the asymptotic performance is measured by the latency, which is the delay in the number of parallel operations, before the result is available. This is in contrast to traditional algorithms, which are assessed by the total number of operations.}. Figure \ref{fig:bitonic_sort} shows an example of how the sorting is performed when $n$ is a power of 2. A bitonic sorting treats the list as bitonic subsets that gets merged over $\log_2(n)$ major rounds, each of which consists of up to $\log_2(n)$ minor rounds to preserve the bitonic property.
After $\log_2(n)$ major rounds, the set is the first half of a bitonic set, which implies that it is sorted. 
% To perform this sorting most efficiently, we spawn a single kernel to sort across all states and assign each block with $\lceil n / 2 \rceil_2$ threads to one source state to sort destination states with non-zero probability, all in shared memory for maximizing performance. $n$ is the maximum number of destination states across all source states.

% \FM{$\lceil \cdot \rceil_2$ means round up to nearest power of two.}

\begin{figure}
    \centering
    \includegraphics[width=0.8\linewidth]{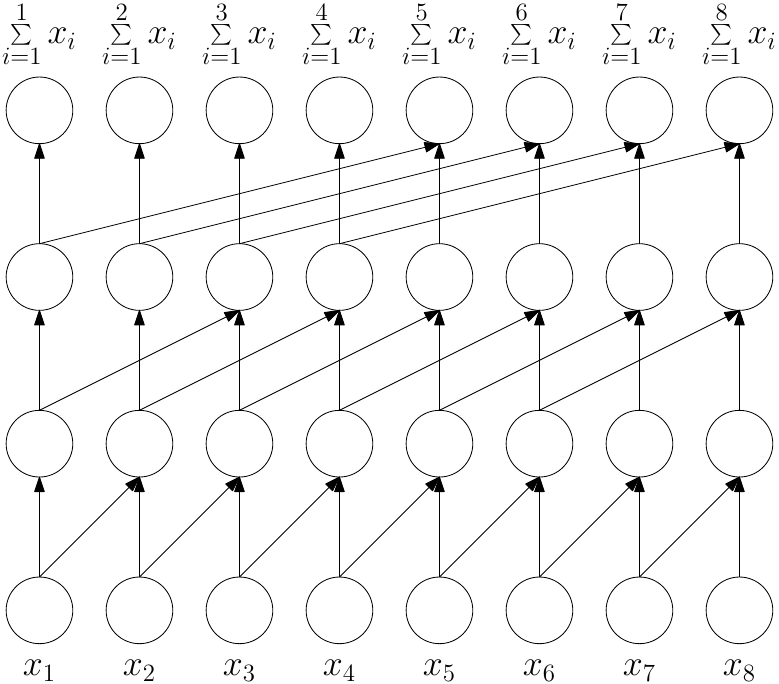}
    \caption{Cumulative sum as a tree-reduction. This tree reduction has a latency of $O(\log_2(n))$ as opposed to $O(n)$ of a traditional iterative algorithm.}
    \label{fig:tree_reduction_cumulative_sum}
\end{figure}
		
The second phase consists of assigning probabilities within the feasible probability intervals to give the most probability mass possible to states early in the ordering. 
% Say that the ordering is in increasing order of $V$, then giving most probability mass to states with smallest $V$ is equivalent to finding a lower bound.
To perform this assignment step in parallel, we have to parallelize the sequential algorithm below.
\begin{lstlisting}[language=JuliaLocal, style=julia]
function omaximization(ordering, lower, upper)
    p = copy(lower)
    rem = 1 - sum(lower)
    gap = upper - lower

    for o in ordering
        if gap[o] < rem
            p[o] += gap[o]
            rem -= gap[o]
        else
            p[o] += rem
            break;
        end
    end

    return p
end
\end{lstlisting}
While seemingly sequential in nature, the method implicitly implements a cumulative summation procedure where the summation is according to the ordering previously found.
% \LL{Be more precise. This is vague. Does it reduce to it? }\FM{The cumulative sum is a subprocedure to the O-maximization. The identification that the original description of O-maximization contains an implicit cumulative sum is what's new here, IMO.}
Consequently, we can employ the parallel algorithm based on tree reduction developed in \citep{ladner1980parallel} (see Figure \ref{fig:tree_reduction_cumulative_sum}), which has latency $O(\log_2(n))$.
%There exists a known method for computing a cumulative sum in parallel using a tree reduction (see Figure \ref{fig:tree_reduction_cumulative_sum}), which has latency $O(\log_2(n))$ (\cite{ladner1980parallel}). 
In particular, for O-maximization, the cumulative sum is computed over the gap between the upper and lower bounds according to the ordering. Thus, each element will have the sum of the gaps of the states up to and including itself in the ordering. Then each iteration in O-maximization is independent and thus can be performed in parallel, that is, with a latency of $O(1)$.

\begin{lstlisting}[language=JuliaLocal, style=julia]
function omaximization(ordering, lower, upper)
    p = copy(lower)
    rem = 1 - sum(lower)
    gap = upper - lower

    # Ordered cumulative sum of gaps
    cumgap = cumulative_sum(gap[ordering])

    for (i, o) in enumerate(ordering)
        rem_state = max(rem - cumgap[i] + gap[o], 0)
        if gap[o] < rem_state
            p[o] += gap[o]
        else
            p[o] += rem_state
        end
    end

    return p
end
\end{lstlisting}

\section{Computational studies}
\label{sec:experiments}
In order to show the effectiveness of \IMDPjl, we compare it against bmdp-tool and PRISM, which, to the best of our knowledge, are the only existing tools that support value iteration for \glspl{imdp}.
%Though we will not report the details here, we have compared the resulting values against both tools to ensure the correctness of our tool. 
\ifbool{inproceedings}{
% iftrue
We benchmark the tools on 35 models that are taken from the literature and include abstractions of linear and non-linear systems, including neural networks, with number of transition probabilities ranging from few tens to tens of millions.
}{
% iffalse
We benchmark the tools on 35 models, whose details can be found in Table \ref{tab:experiment_imdp_models} in Appendix \ref{app:benchmarking_models}. The models are taken from the literature and include abstractions of linear and non-linear systems, including neural networks, with number of transition probabilities ranging from few tens to tens of millions.
}

%originate from multiple sources - one is included with bmdp-tool, three are built as product \glspl{imdp} to prove LTLf properties, and 18 are abstractions of Neural Network Dynamic Models, and the remaining 13 are abstractions of linear and non-linear systems.  
For a fair comparison, for all models we run a (maximum pessimistic) finite time reachability query (200 time steps).
All experiments were conducted on a computer with 16GB RAM, an Intel I7-6700K CPU (4 cores, 8 hyperthreads), and an NVIDIA GTX1060 6GB VRAM GPU.
For each tool, we measure the computation time only, and not the time it takes to load each file.
% As the measurement is only for the computation, none of the tools will explicitly yield to the operating system. 
% Note that PRISM is written in Java and \IMDPjl is written in Julia, both of which are garbage-collected languages. Therefore, in particular for smaller models, the measurements may exhibit more variance compared to bmdp-tool, which is written in C++.

\begin{figure}
    \centering
    \begin{subfigure}[b]{\linewidth}
        \centering
        \includegraphics[width=\textwidth]{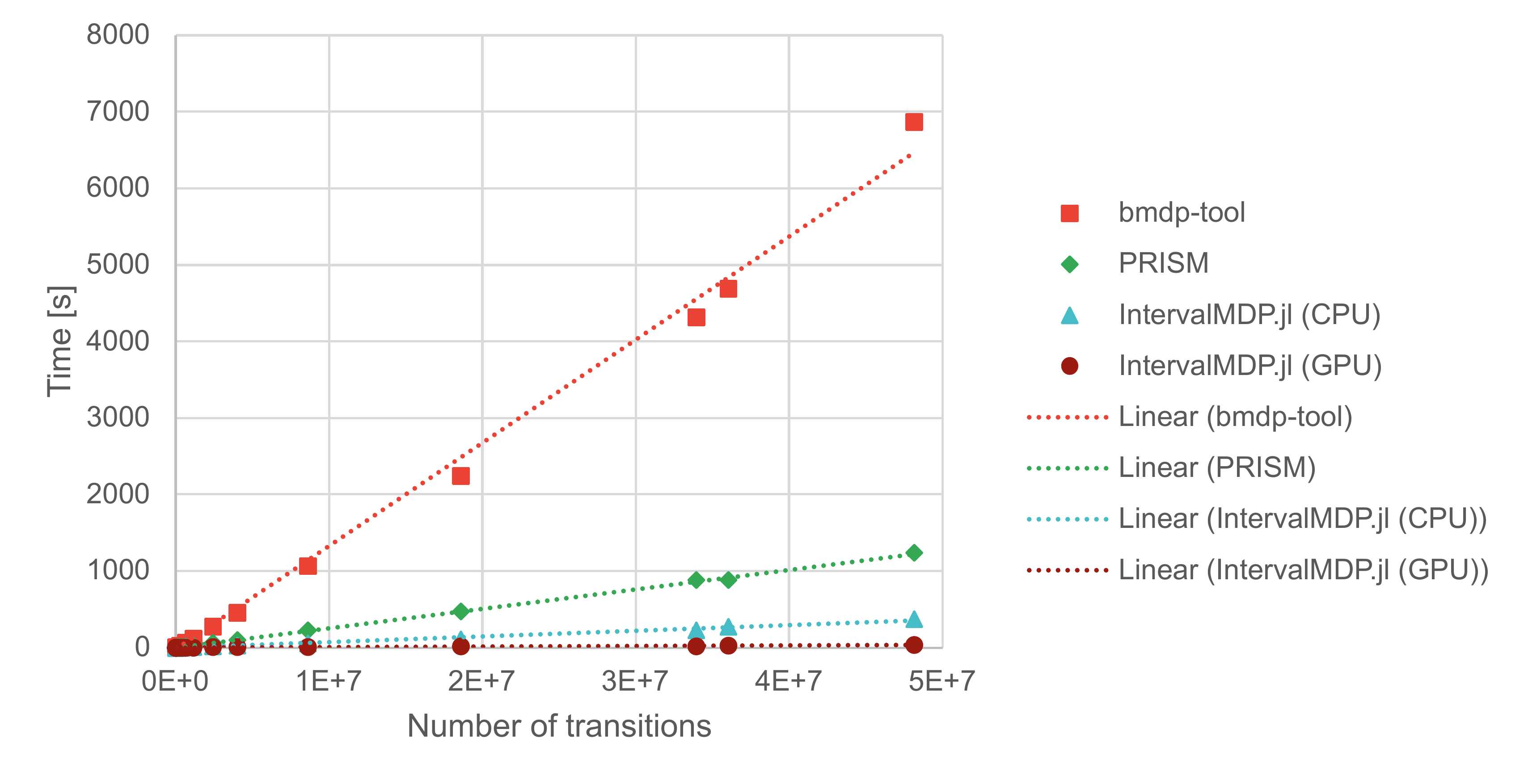}
        \caption{Linear scale with trendline}
        \label{fig:computation_time_linear}
    \end{subfigure}
    \par\bigskip
    \begin{subfigure}[b]{\linewidth}
        \centering
        \includegraphics[width=\textwidth]{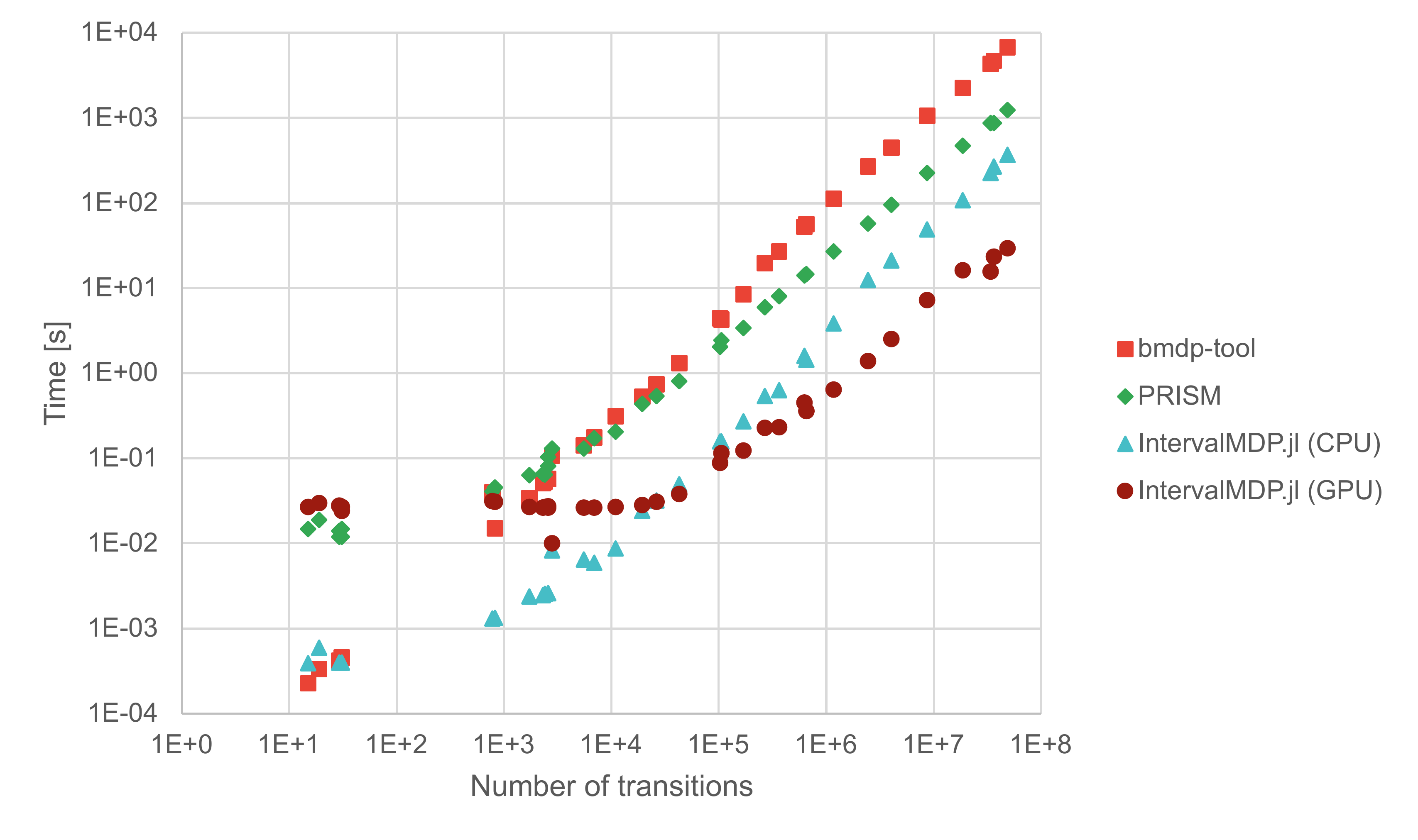}
        \caption{Log-log scale}
        \label{fig:computation_time_log}
    \end{subfigure}
    \caption{Computation time of \IMDPjl compared against bmdp-tool and PRISM on 35 \glspl{imdp} of varying sizes.}
    \label{fig:computation_time_with_number_of_transitions}
\end{figure}

The obtained results are shown in Figure~\ref{fig:computation_time_with_number_of_transitions}
as plots of computation time as a function of the number of transitions in the \gls{imdp}. 
% First, comparing bmdp-tool and PRISM, the log-log plot in Figure \ref{fig:computation_time_with_number_of_transitions} shows that bmdp-tool is faster on the smallest models, but PRISM is faster on larger models. 
In Figure~\ref{fig:computation_time_linear}, we see that \IMDPjl substantially outperforms the other tools in terms of computation time. For instance, computing the query for the largest model takes bmdp-tool $6865$s and PRISM $1235$s, while \IMDPjl takes $372$s and $30$s for the CPU and GPU implementation respectively. This is a speed-up of 228$\times$ and 41$\times$ of the GPU implementation of \IMDPjl relative to bmdp-tool and PRISM respectively. 

The log-log plot in Figure~\ref{fig:computation_time_log} highlights the performance differences on the smaller models. Specifically, the bmdp-tool and the CPU implementation of \IMDPjl are faster than PRISM on the smallest models, with a speed-up of 30-60$\times$.
We conjecture that the better scaling of PRISM compared to bmdp-tool is due to using a dictionary of non-zero probability destinations for each source-action pair stored in a list, rather than a dictionary keyed by source/action/destination triplets.
\IMDPjl, on the other hand, does not store any dictionary. Instead, we track and store the indices for source, action, and destination in the \gls{csc}-format sequentially, allowing better caching when accessing probabilities.
In Figure \ref{fig:computation_time_log}, we also see that the GPU implementation of \IMDPjl has overhead that is dominant for smaller models, while for larger models is consistently orders of magnitude faster than the other implementations. The cut-off is roughly at 25000 transitions in the \gls{imdp}.

A common bottleneck for \gls{imdp} tools is memory consumption, which is only exacerbated on a GPU, as they generally have less memory available. 
However, due to the \gls{csc}-format with \texttt{Float64} values and \texttt{Int32} indices, \IMDPjl generally requires less memory compared to PRISM and bmdp-tool. For example, to run value iteration on the largest model\xspace\ifbool{inproceedings}{}{(i.e., \texttt{pimdp\_2} in Table \ref{tab:experiment_imdp_models} in Appendix \ref{app:benchmarking_models})}, \IMDPjl requires $4.88$GB of memory. In contrast, PRISM uses $6.32$GB of memory and bmdp-tool uses $5.38$GB to run value iteration on the same problem. This is a 23\% reduction relative to PRISM and 9\% relative to bmdp-tool, which is including the Julia runtime.
%Therefore, we will here describe an analytical formula for the memory requirements of our tool.
%Assume that the model is stored in \gls{csc}-format with \texttt{Float64} values and \texttt{Int32} indices. Then to run value iteration on the GPU using our tool takes $3 \cdot ((8 + 4)n + 4(m + 1)) + (4n + 4(m + 1))$ bytes of memory where $n$ is the number of transitions and $m$ is the number of choices, which is the number of state multiplied by the number of actions. The first term is the memory required to store the lower bound, the gap or upper bound, and the resulting probability assignment. The second term is the memory required to store the ordering of states for each state/action pair. 

\section{Conclusion}
We introduced \IMDPjl, a new software package for robust value iteration of IMDPs that employs parallel algorithms and takes advantage of CUDA-accelerated hardware.
Computational experiments showed speed-ups of $50$-$200\times$ relative to state-of-the-art.
\IMDPjl is currently limited to reachability specifications along with reward optimization. We plan to extend it to support Linear Temporal Logic \citep{Baier2008}, via the product construction \citep{10.1145/1967701.1967715}, and Probabilistic Computation Tree Logic \citep{Hansson1994-ma}, via parse tree construction \citep{7029024}. 

%% There are a number of predefined theorem-like environments in
%% ifacconf.cls:
%%
%% \begin{thm} ... \end{thm}            % Theorem
%% \begin{lem} ... \end{lem}            % Lemma
%% \begin{claim} ... \end{claim}        % Claim
%% \begin{conj} ... \end{conj}          % Conjecture
%% \begin{cor} ... \end{cor}            % Corollary
%% \begin{fact} ... \end{fact}          % Fact
%% \begin{hypo} ... \end{hypo}          % Hypothesis
%% \begin{prop} ... \end{prop}          % Proposition
%% \begin{crit} ... \end{crit}          % Criterion

% \begin{ack}
% Place acknowledgements here.
% \end{ack}

\bibliography{ifacconf}             % bib file to produce the bibliography
                                                     % with bibtex (preferred)
                                                   
%\begin{thebibliography}{xx}  % you can also add the bibliography by hand

%\bibitem[Able(1956)]{Abl:56}
%B.C. Able.
%\newblock Nucleic acid content of microscope.
%\newblock \emph{Nature}, 135:\penalty0 7--9, 1956.

%\bibitem[Able et~al.(1954)Able, Tagg, and Rush]{AbTaRu:54}
%B.C. Able, R.A. Tagg, and M.~Rush.
%\newblock Enzyme-catalyzed cellular transanimations.
%\newblock In A.F. Round, editor, \emph{Advances in Enzymology}, volume~2, pages
%  125--247. Academic Press, New York, 3rd edition, 1954.

%\bibitem[Keohane(1958)]{Keo:58}
%\newblock \emph{Power and Interdependence: World Politics in Transitions}.
%\newblock Little, Brown \& Co., Boston, 1958.

%\bibitem[Powers(1985)]{Pow:85}
%\newblock Is there a way out?
%\newblock \emph{Harpers}, pages 35--47, June 1985.

%\bibitem[Soukhanov(1992)]{Heritage:92}
%\newblock \emph{{The American Heritage. Dictionary of the American Language}}.
%\newblock Houghton Mifflin Company, 1992.

%\end{thebibliography}

\ifinproceedings
%iftrue
\else
%iffalse

\appendix
\section{Detailed description of storage formats}    % Each appendix must have a short title.
\label{sec:appendix_storage_formats}

\subsection{\IMDPjl}
The new format proposed in this paper uses netCDF, which is based on HDF5 underlying, to store transition probabilities, and a JSON file to store the specification. Transition probabilities are stored in \gls{csc}-format, which is unfortunately not natively stored in netCDF, nor any widely available format.
Therefore, we store the following attributes and variables in the netCDF file:~\\[0.5em]
\begin{tabularx}{\linewidth}{>{\hsize=0.5\hsize\arraybackslash}X>{\hsize=1.5\hsize\arraybackslash\linewidth=\hsize}X}
     \toprule
     Global attributes & 
        \begin{itemize}[leftmargin=5pt, labelindent=0pt, itemindent=0pt, before=\vspace{-0.75\baselineskip}, after=\vspace{-0.5\baselineskip}]
            \item \texttt{num\_states}
            \item \texttt{model} (either \texttt{imc} or \texttt{imdp})
            \item \texttt{format} (assert \texttt{sparse\_csc})
            \item \texttt{rows} (assert \texttt{to})
            \item \texttt{cols} (assert \texttt{from} if model is \texttt{imc} and \texttt{from/action} if model is \texttt{imdp})
        \end{itemize} \\
    Variables &
        \begin{itemize}[leftmargin=5pt, labelindent=0pt, itemindent=0pt, before=\vspace{-0.75\baselineskip}, after=\vspace{-0.5\baselineskip}]
            \item \texttt{lower\_colptr} (integer)
            \item \texttt{lower\_rowval} (integer)
            \item \texttt{lower\_nzval} (floating point)
            \item \texttt{upper\_colptr} (integer)
            \item \texttt{upper\_rowval} (integer)
            \item \texttt{upper\_nzval} (floating point)
            \item \texttt{stateptr} (integer, only for \texttt{imdp})
            \item \texttt{action\_vals} (any netCDF supported type, only for \texttt{imdp})
        \end{itemize}\\
     \bottomrule
\end{tabularx}

We store the specification in a JSON format where the structure depends on the type of specification.
For a reachability-like specification, the specification is the following format
\begin{lstlisting}[language=json]
{
    "property": {
        "type": <"reachability"|"reach-avoid">,
        "infinite_time": <true|false>,
        "time_horizon": <positive int>,
        "eps": <positive float>,
        "reach": [<state_index:positive int>],
        "avoid": [<state_index:positive int>]
    },
    "satisfaction_mode": <"pessimistic"|"optimistic">,
    "strategy_mode": <"minimize"|"maximize">
}
\end{lstlisting}
For a finite horizon property, \texttt{eps} is excluded, and similarly for an infinite horizon property, \texttt{time\_horizon} is excluded. 
For a proper reachability property, the \texttt{avoid}-field is excluded.

If we instead want to optimize a reward, the format is the following
\begin{lstlisting}[language=json]
{
    "property": {
        "type": "reward",
        "infinite_time": <true|false>,
        "time_horizon": <positive int>,
        "eps": <positive float>,
        "reward": [<reward_per_state_index:float>]
        "discount" <float:0-1>
    },
    "satisfaction_mode": <"pessimistic"|"optimistic">,
    "strategy_mode": <"minimize"|"maximize">
}
\end{lstlisting}

\section{Benchmarking models}
\label{app:benchmarking_models}

\begin{table*}[b]
  \centering
  \caption{List of the 35 models that was used in the empirical study (see Section \ref{sec:experiments}) and their properties.}
  \label{tab:experiment_imdp_models}
  \begin{tabular}{lccc}
    \toprule
    Model & Number of states & Number of actions & Number of transitions \\\midrule
    \texttt{pimdp\_0} & 21174 & 3 & 33908939 \\
    \texttt{pimdp\_1} & 32336 & 3 & 36019324 \\
    \texttt{pimdp\_2} & 42634 & 3 & 48106480 \\
    \texttt{multiObj\_robotIMDP} & 207 & 4 & 2784 \\
    \texttt{linear\_5\_states\_0.9\_f\_0.01\_sigma} & 6 & 1 & 15 \\
    \texttt{linear\_5\_states\_0.9\_f\_0.05\_sigma} & 6 & 1 & 29 \\
    \texttt{linear\_5\_states\_0.9\_f\_0.1\_sigma} & 6 & 1 & 31 \\
    \texttt{linear\_5\_states\_1.05\_f\_0.01\_sigma} & 6 & 1 & 19 \\
    \texttt{linear\_5\_states\_1.05\_f\_0.05\_sigma} & 6 & 1 & 29 \\
    \texttt{linear\_5\_states\_1.05\_f\_0.1\_sigma} & 6 & 1 & 31 \\
    \texttt{linear\_50\_states\_0.9\_f\_0.01\_sigma} & 51 & 1 & 819 \\
    \texttt{linear\_50\_states\_0.9\_f\_0.05\_sigma} & 51 & 1 & 2379 \\
    \texttt{linear\_50\_states\_0.9\_f\_0.1\_sigma} & 51 & 1 & 2551 \\
    \texttt{linear\_50\_states\_1.05\_f\_0.01\_sigma} & 51 & 1 & 773 \\
    \texttt{linear\_50\_states\_1.05\_f\_0.05\_sigma} & 51 & 1 & 2299 \\
    \texttt{linear\_50\_states\_1.05\_f\_0.1\_sigma} & 51 & 1 & 2551 \\
    \texttt{pendulum\_1\_layer\_120\_states\_0.01\_sigma} & 121 & 1 & 1709 \\
    \texttt{pendulum\_1\_layer\_120\_states\_0.05\_sigma} & 121 & 1 & 6877 \\
    \texttt{pendulum\_1\_layer\_120\_states\_0.1\_sigma} & 121 & 1 & 10839 \\
    \texttt{pendulum\_1\_layer\_240\_states\_0.01\_sigma} & 241 & 1 & 5489 \\
    \texttt{pendulum\_1\_layer\_240\_states\_0.05\_sigma} & 241 & 1 & 26321 \\
    \texttt{pendulum\_1\_layer\_240\_states\_0.1\_sigma} & 241 & 1 & 42480 \\
    \texttt{pendulum\_1\_layer\_480\_states\_0.01\_sigma} & 481 & 1 & 19323 \\
    \texttt{pendulum\_1\_layer\_480\_states\_0.05\_sigma} & 481 & 1 & 102213 \\
    \texttt{pendulum\_1\_layer\_480\_states\_0.1\_sigma} & 481 & 1 & 167398 \\
    \texttt{cartpole\_1\_layer\_960\_states\_0.01\_sigma} & 961 & 1 & 105047 \\
    \texttt{cartpole\_1\_layer\_960\_states\_0.05\_sigma} & 961 & 1 & 361621 \\
    \texttt{cartpole\_1\_layer\_960\_states\_0.1\_sigma} & 961 & 1 & 647640 \\
    \texttt{cartpole\_1\_layer\_1920\_states\_0.01\_sigma} & 1921 & 1 & 266855 \\
    \texttt{cartpole\_1\_layer\_1920\_states\_0.05\_sigma} & 1921 & 1 & 1159587 \\
    \texttt{cartpole\_1\_layer\_1920\_states\_0.1\_sigma} & 1921 & 1 & 2432256 \\
    \texttt{cartpole\_1\_layer\_3840\_states\_0.01\_sigma} & 3841 & 1 & 620296 \\
    \texttt{cartpole\_1\_layer\_3840\_states\_0.05\_sigma} & 3841 & 1 & 4014802 \\
    \texttt{cartpole\_1\_layer\_3840\_states\_0.1\_sigma} & 3841 & 1 & 8614913 \\
    \texttt{harrier\_25920\_states\_[0.05,0.05,0.02,0.01,0.01,0.01]\_sigma} & 25921 & 1 & 18574307 \\ \bottomrule
  \end{tabular}
\end{table*}
\fi

\end{document}

%% file: macros.tex
\newcommand{\IMDPjl}{\texttt{IntervalMDP.jl}\xspace}

\makeatletter
\newsavebox\myboxA
\newsavebox\myboxB
\newlength\mylenA
\newcommand*\xoverline[2][0.75]{%
    \sbox{\myboxA}{$\m@th#2$}%
    \setbox\myboxB\null% Phantom box
    \ht\myboxB=\ht\myboxA%
    \dp\myboxB=\dp\myboxA%
    \wd\myboxB=#1\wd\myboxA% Scale phantom
    \sbox\myboxB{$\m@th\overline{\copy\myboxB}$}%  Overlined phantom
    \setlength\mylenA{\the\wd\myboxA}%   calc width diff
    \addtolength\mylenA{-\the\wd\myboxB}%
    \ifdim\wd\myboxB<\wd\myboxA%
       \rlap{\hskip 0.5\mylenA\usebox\myboxB}{\usebox\myboxA}%
    \else
        \hskip -0.5\mylenA\rlap{\usebox\myboxA}{\hskip 0.5\mylenA\usebox\myboxB}%
    \fi}
\makeatother

\DeclareMathOperator*{\opta}{opt^\pi}
\DeclareMathOperator*{\opte}{opt^\eta}

%% file: acronyms.tex
\newacronym[longplural=Interval Markov Decision Processes]{imdp}{IMDP}{Interval Markov Decision Process}
\newacronym[longplural=Markov Decision Processes]{mdp}{MDP}{Markov Decision Process}
\newacronym{gpu}{GPU}{General-Purpose Graphics Processing Unit}
\newacronym{csc}{CSC}{Compressed Sparse Column}
\newacronym{simd}{SIMD}{Single Instruction Multiple Data}
\newacronym{simt}{SIMT}{Single Instruction Multiple Thread}
\newacronym{api}{API}{Application Programming Interface}

%% file: root.bbl
\begin{thebibliography}{18}
\providecommand{\natexlab}[1]{#1}
\providecommand{\url}[1]{\texttt{#1}}
\providecommand{\urlprefix}{URL }
\expandafter\ifx\csname urlstyle\endcsname\relax
  \providecommand{\doi}[1]{doi:\discretionary{}{}{}#1}\else
  \providecommand{\doi}{doi:\discretionary{}{}{}\begingroup \urlstyle{rm}\Url}\fi

\bibitem[{Abate et~al.(2011)Abate, Katoen, and Mereacre}]{10.1145/1967701.1967715}
Abate, A., Katoen, J.P., and Mereacre, A. (2011).
\newblock Quantitative automata model checking of autonomous stochastic hybrid systems.
\newblock In \emph{Proceedings of the 14th international conference on Hybrid systems: computation and control}, 83--92.

\bibitem[{Baier and Katoen(2008)}]{Baier2008}
Baier, C. and Katoen, J.P. (2008).
\newblock \emph{Principles of Model Checking}.
\newblock The MIT Press.

\bibitem[{Bezanson et~al.(2012)Bezanson, Karpinski, Shah, and Edelman}]{bezanson2012julia}
Bezanson, J., Karpinski, S., Shah, V.B., and Edelman, A. (2012).
\newblock Julia: A fast dynamic language for technical computing.
\newblock \emph{arXiv preprint arXiv:1209.5145}.

\bibitem[{Bulu{\c{c}} et~al.(2009)Bulu{\c{c}}, Fineman, Frigo, Gilbert, and Leiserson}]{10.1145/1583991.1584053}
Bulu{\c{c}}, A., Fineman, J.T., Frigo, M., Gilbert, J.R., and Leiserson, C.E. (2009).
\newblock Parallel sparse matrix-vector and matrix-transpose-vector multiplication using compressed sparse blocks.
\newblock In \emph{Proceedings of the twenty-first annual symposium on Parallelism in algorithms and architectures}, 233--244.

\bibitem[{Cauchi et~al.(2019)Cauchi, Laurenti, Lahijanian, Abate, Kwiatkowska, and Cardelli}]{cauchi2019efficiency}
Cauchi, N., Laurenti, L., Lahijanian, M., Abate, A., Kwiatkowska, M., and Cardelli, L. (2019).
\newblock Efficiency through uncertainty: Scalable formal synthesis for stochastic hybrid systems.
\newblock In \emph{Proceedings of the 22nd ACM international conference on hybrid systems: computation and control}, 240--251.

\bibitem[{Delimpaltadakis et~al.(2023)Delimpaltadakis, Lahijanian, Mazo~Jr, and Laurenti}]{delimpaltadakis2023interval}
Delimpaltadakis, G., Lahijanian, M., Mazo~Jr, M., and Laurenti, L. (2023).
\newblock Interval markov decision processes with continuous action-spaces.
\newblock In \emph{Proceedings of the 26th ACM International Conference on Hybrid Systems: Computation and Control}, 1--10.

\bibitem[{Givan et~al.(2000)Givan, Leach, and Dean}]{givan2000bounded}
Givan, R., Leach, S., and Dean, T. (2000).
\newblock Bounded-parameter markov decision processes.
\newblock \emph{Artificial Intelligence}, 122(1-2), 71--109.

\bibitem[{Haddad and Monmege(2018)}]{HADDAD2018111}
Haddad, S. and Monmege, B. (2018).
\newblock Interval iteration algorithm for mdps and imdps.
\newblock \emph{Theoretical Computer Science}, 735, 111--131.

\bibitem[{Hansson and Jonsson(1994)}]{Hansson1994-ma}
Hansson, H. and Jonsson, B. (1994).
\newblock A logic for reasoning about time and reliability.
\newblock \emph{Formal aspects of computing}, 6, 512--535.

\bibitem[{Kwiatkowska et~al.(2011)Kwiatkowska, Norman, and Parker}]{kwiatkowska2011prism}
Kwiatkowska, M., Norman, G., and Parker, D. (2011).
\newblock Prism 4.0: Verification of probabilistic real-time systems.
\newblock In \emph{Computer Aided Verification: 23rd International Conference}, 585--591. Springer.

\bibitem[{Ladner and Fischer(1980)}]{ladner1980parallel}
Ladner, R.E. and Fischer, M.J. (1980).
\newblock Parallel prefix computation.
\newblock \emph{Journal of the ACM (JACM)}, 27(4), 831--838.

\bibitem[{Lahijanian(2021)}]{morteza_lahijanian_bmdp_tool}
Lahijanian, M. (2021).
\newblock {bmdp-tool}.
\newblock \urlprefix\url{https://github.com/aria-systems-group/bmdp-tool}.
\newblock Standalone software.

\bibitem[{Lahijanian et~al.(2015)Lahijanian, Andersson, and Belta}]{7029024}
Lahijanian, M., Andersson, S.B., and Belta, C. (2015).
\newblock Formal verification and synthesis for discrete-time stochastic systems.
\newblock \emph{IEEE Transactions on Automatic Control}, 60(8), 2031--2045.

\bibitem[{Laurenti and Lahijanian(2023)}]{laurenti2023unifying}
Laurenti, L. and Lahijanian, M. (2023).
\newblock Unifying safety approaches for stochastic systems: From barrier functions to uncertain abstractions via dynamic programming.
\newblock \emph{arXiv preprint arXiv:2310.01802}.

\bibitem[{Lavaei et~al.(2022)Lavaei, Soudjani, Abate, and Zamani}]{lavaei2022automated}
Lavaei, A., Soudjani, S., Abate, A., and Zamani, M. (2022).
\newblock Automated verification and synthesis of stochastic hybrid systems: A survey.
\newblock \emph{Automatica}, 146, 110617.

\bibitem[{{NVIDIA Corporation}(2023)}]{cuda}
{NVIDIA Corporation} (2023).
\newblock {NVIDIA CUDA C++} programming guide.
\newblock Version 12.3.

\bibitem[{Peters et~al.(2010)Peters, Schulz-Hildebrandt, and Luttenberger}]{10.1007/978-3-642-14390-8_42}
Peters, H., Schulz-Hildebrandt, O., and Luttenberger, N. (2010).
\newblock Fast in-place sorting with cuda based on bitonic sort.
\newblock In \emph{Parallel Processing and Applied Mathematics: 8th International Conference}, 403--410. Springer.

\bibitem[{Rew et~al.(2006)Rew, Hartnett, Caron et~al.}]{rew2006netcdf}
Rew, R., Hartnett, E., Caron, J., et~al. (2006).
\newblock Netcdf-4: Software implementing an enhanced data model for the geosciences.
\newblock In \emph{22nd International Conference on Interactive Information Processing Systems for Meteorology, Oceanograph, and Hydrology}.

\end{thebibliography}
